# An Encrypted Trust-Based Routing Protocol

Youssef Gahi[1], Mouhcine Guennoun[2], Zouhair Guennoun[1], Khalil El-Khatib[2]
[1]Laboratoire d'Electronique et de Communications – LEC, Ecole Mohammadia d'Ingénieurs – EMI,
Université Mohammed V-Agdal – UM5A. BP 765, avenue Ibn Sina Agdal 10000, Rabat, Morocco.
[2]University of Ontario Institute of Technology, 2000 Simcoe Street North,
Oshawa, Ontario, Canada. L1H 7K4
youssef.gahi@gmail.com, mouhcine.guennoun@uoit.ca, zouhair@emi.ac.ma, khalil.el-khatib@uoit.ca

*Abstract*—The interest in trust-based routing protocols has grown with the advancements achieved in ad-hoc wireless networks. However, regardless of the many security approaches and trust metrics available, trust-based routing still faces some security challenges, especially with respect to privacy. In this paper, we propose a novel trust-based routing protocol based on a fully homomorphic encryption scheme. The new protocol allows nodes, which collaborate in a dynamic environment, to evaluate their knowledge on the trustworthiness of specific routes and securely share this knowledge.

## I. INTRODUCTION

An ad-hoc wireless network is a set of dynamic nodes connected to each other without any physical liaison or pre-conceived infrastructure. This technology, which is suitable for dynamic topologies, has attracted a valuable attention from the research community. Researchers have discussed the various aspects of realizing a strong, dynamic architecture that matches its counterpart based on some configured parameters. In this context, routing protocols receive the utmost priority since they deal with the paths traversed by packets in dynamic networks.

Routing in ad-hoc wireless networks relies on the collaboration of nodes to forward a packet from its source to a targeted destination. However, if insufficient security measures are implemented, the forwarding functionality may be easily compromised by selfish/malicious nodes that mislead legitimate nodes to forward packets away from their designated destination. Clearly, nodes tend to collaborate with only legitimate (honest) nodes to achieve high rates of packet delivery. In this context, an interesting approach has been proposed [1] to avoid forwarding packets over untrustworthy paths. With this approach, each node evaluates its neighbors to elect the ones that it will collaborate with. A trust value is assigned to each node based on its cooperation. This value is updated continuously. The trust values are then used to define a trust route. This approach suffers from a major limitation. Basically, no special measures are taken to protect the trust values generated by the evaluating node. This is a sensitive piece of information that should be hidden from malicious nodes. Therefore, we focus in this work on providing protective measure that can preserve the privacy of nodes without degrading the robustness in performance. We utilize the concept of Fully Homomorphic Encryption and Multi-hop Homomorphic to achieve our goal.

The Fully Homomorphic Encryption (FHE) is an efficient concept that can enable the manipulation of encrypted data in a blind fashion. FHE schemes allow performing algebraic computations such as addition and multiplication, unlike other schemes which support only a single type of operations (addition or multiplication) [12-15]. This capability provides a high level of flexibility and makes it possible to publish encrypted data as well as delegate encrypted data processing to third parties. This proves beneficial for packets routing, since each node can perform its trust circuit and then communicate, to its neighbors, the resulting trust evaluation in an encrypted format. This way, nodes can effectively protect their privacy. It is worth mentioning, however, that the FHE concept is bounded by the number of circuits that can participate in a certain computation. This is can be understood by noticing that in the output of a circuit, the resulting ciphertext is no fresher as it was in entry, and therefore, FHE properties might be no more valid. To overcome this limitation, we use a mechanism called the "Adapter", which implements a multi-hop concept to preserve the FHE properties, even when multiple hops are involved. Such a mechanism allows us to develop an efficient and secure trust-based routing protocol for ad-hoc wireless networks.

The remainder of this paper is organized as follows. In Section II we review the related work to the area of trust-based routing protocols. Section III provides an overview of routing techniques as well as their limitations. Section IV presents our novel trust-based routing protocol. Section V details the solution and describes the designed protocol. Section VI we study the performance of our new protocol. Finally, Section VII concludes our work and provides future research directions.

## II. RELATED WORK

In this section, we briefly describe other works that aim at defining trust mechanisms or designing secure routing protocols in ad-hoc wireless networks.

Pirzada et al. define a trust mechanism [1] for a reactive routing protocol. They propose a model in which each node in an ad-hoc wireless environment maintains an evaluation procedure to reward or punish nodes in future collaborations. The evaluation generates trust values, which are based on transactions' history and the forwarding quality, that are shared with other nodes to choose a trust path. However, the proposal in [1] does not employ any measures to protect the evaluations conducted by the nodes and any malicious node can access these trust values and harm the routing functionality.

Nekkanti and Lee propose in [2] a trust-based adaptive routing protocol that denies nodes which are not affected to access route information about routes' path. Furthermore, this contribution proposes a mechanism that uses different kinds of encryptions that vary according to the nature of nodes and their trust level. These techniques are employed to protect the route privacy and to save energy without affecting the performance.

Chou et al. in [3] adopt a trust-based reputation mechanism for the reactive routing protocol DSR. This contribution proposes a model that detects malicious and selfish nodes in order to prevent them from participating in the routing process.

Sanzgiri et al. in [4] propose a secure routing protocol based on authentication mechanisms.

Zhang et al. in [5] provide a formal study of trust-based routing in ad-hoc wireless networks. The work discusses several aspects related to these protocols, such as correctness, respecting optimal trajectory to construct paths, and distributive of trust branches.

Although the abovementioned contributions provide important discussions that tackle various aspects of trust-based routing protocols, none of them has addressed the support of privacy in these protocols. We focus on this problem in this paper.

## III. TRUST BASED ROUTING PROTOCOL

### A. Routing protocol

Routing protocols define mechanisms by which a router finds the paths over which packets are sent to reach their targeted destination. These protocols define routing rules, manage priorities, and report any changes faced by the nodes. Routing protocols can be categorized as static (pro-active) protocols or dynamic (reactive) protocols. Static routing depends on a pre-configured set of routes that are ready for use by the nodes to deliver packets to their destinations. Moreover, pro-active protocols discover routes on-demand by sending Route Request packets over the network. Dynamic routing, however, constructs the route depending on dynamic protocols that update the routing rules based on the network's changing conditions. Both of these techniques have strengths and weaknesses, depending on the adopted network topology. Since, it is sometimes possible to overlook limitations of the Reactive protocol to achieve exact routing and better administration. Whereas in other architecture, where flexibility and dynamic raise as an important factor, Reactive protocol is the favorite.

### B. Wireless Ad-Hoc Network

A wireless ad-hoc network is constituted by a standalone architecture that is infrastructure-less, with no specific topology pre-defined. The nodes form the network by collaborating dynamically to construct on-demand paths that are traversed by packets. Ad hoc wireless networks find a broad range of applications in environments where minimal configuration and quick deployment are required. These networks are a typical example where reactive routing protocols are highly preferred and effective. However, routing in ad hoc wireless networks faces major challenges in terms of quality and delivery guarantees. This is because of that the nodes in these networks cooperate and voluntarily announce their availability in order for the forwarding of packets to function properly. Therefore, the routing functionality can fail drastically when a malicious node manages to collude with other nodes to control the network, or when some nodes refuse to collaborate to save their energy. As a result, the selection of nodes that participate in the routing process should be done carefully such that a secure and protected environment can be achieved. This proves the importance of developing trust-based routing protocols that select the nodes based on their level of commitment.

### C. Trust-Based Routing

The concept of *trust*-based routing implies that the routing functionality selects the nodes based on their involvement and degree of faithfulness. The *trust* factor can be quantified based on nodes' cooperation history and reputation. This history can be constructed by monitoring nodes' availability, level of successful packet transfer, and preservation of packets' priority. Based on this history, a trust value is assigned to each node by its predecessors, which facilitates the forming of a trusted path to forward packets over. However, available trust-based routing protocols do not take precautions to protect the process of evaluating nodes. This means that we need to hide the criteria used to assess nodes' trustworthiness in order to prevent malicious nodes from compromising the routing functionality. It is the main objective of this paper to protect the privacy of nodes and allow them to share their experiences safely.

## IV. HOMOMORPHIC ENCRYPTION SCHEMES

### A. Fully homomorphic encryption scheme

The concept of Homomorphic encryption allows performing arithmetic operations over encrypted values. Homomorphic encryption schemes have seen major progresses, starting from supporting either addition (using additive homomorphic encryption) or multiplication (using multiplicative homomorphic encryption), and ending with the ability to perform both operations based on a fully homomorphic encryption scheme (proposed by Gentry [11]). Gentry's scheme achieves effective data protection since a client can publish the data in an encrypted format and delegate to other parties the processing of that data without decrypting them. This scheme can still guarantee a successful decryption of the data after applying several arithmetic operations over its encrypted format.

Gentry's scheme is a polynomial public key encryption scheme which can be designed as $\varepsilon_{pk}(m) = c = m + 2r + pk * Q$, where m is a bit value, r is a random integer, Q is the security number and pk is the public key. The combination of these elements results in c, which is an encrypted integer form of m. The decryption is simply retrieved by computing $m = c \% sk \% 2$, where sk is the secret key. The length of the ciphertext is based on a security parameter called $\lambda$, which is employed to define pk as an odd $\lambda^3$-bit number, besides r (a $\lambda$-bit number) and Q (a $\lambda^2$-bit number). Consequently, each encrypted bit is determined in at most $(1 + \text{Log}_{10}(P))$ decimal digits, where P is a $\lambda^5$-bit number. The Homomorphic circuit can then accept the encrypted values and successfully apply a set of operations over them. These operations generate new encrypted values that can be decrypted only by the party owning the secret key sk, see Figure1.

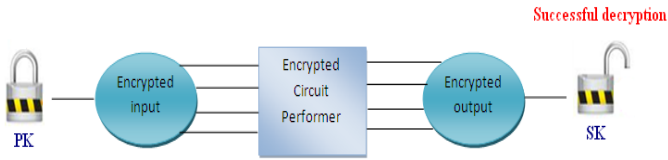

Figure 1. Processing ecnrypted data.

Although the scheme suffers from some drawbacks, like supporting only a limited number of operations and being time-consuming, it is a novel and ingenious approach of cryptography that can provide great added values to security support. To understand the full potential of Gentry's scheme it is important to establish a set of important concepts, like double encryption, homomorphism using different keys, and multi-hop encryption. The latter is the main focus of our study. Multi-hop encryption enables the processing of encrypted entries using different sequenced circuits, see Figure 2. The output ciphertext of this circuit, however, is not as fresh as it was at the entry, and there are no guarantees that a correct decrypted data can be generated from this ciphertext. Therefore, we need to propose an alternative scheme that can perform a fully homomorphic encryption in a multi-hop environment.

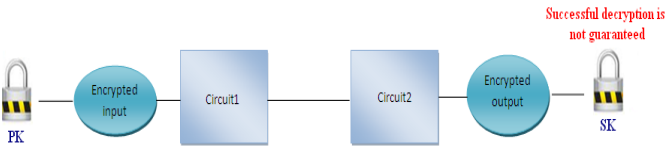

Figure 2. Multi-hop encryption.

### B. Multi-hop Fully Homomorphic Encryption Scheme

Gentry et al. [6] discussed the aforementioned problem in the case of double encryption, and proposed a solution based on the Secure Function Evaluation [7] and Yao's Garbled circuit [8]. This solution consists of extending the principle of two-party computation to support the i-hop homomorphic encryption scheme for a fixed integer i. Recall that two-party computation allows performing a circuit with Heterogeneous inputs. Namely, a server Bob who owns a circuit $C$ can combine his input $x$ (known) with another input y (unknown) to execute his circuit for these values, while ignoring the exact value of $y$ and without revealing the value of its input x. This technique arises as a typical solution of the Millionaires' problem that aims at deciding who is richest without revealing the wealth of both participants. On the other hand, the fact that Gentry relies on such a mechanism to fix the i-hop theory is discouraging as far as it is promising. This is because of that the solution inherits all the drawbacks of Yao's garbled circuits, such as pre-calculating the whole circuit, transferring the whole truth table from one circuit to another, and being time-consuming. The advantage of Gentry's scheme, however, is that it is suitable for a topology that uses multiple circuits with different public keys.

In our contribution, we have a slight difference compared to Gentry's scheme in [6]. We utilize only one public key, and all circuits conduct their computation based on the same shared key. We therefore propose a simplified technique to preserve the Fully Homomorphic Encryption properties in a multi-hop topology.

The main difficulty experienced in an environment containing multiple circuits is that a module $C_1$ acts on the initial fresh ciphertext sent by the client, and outputs a processed one. The latter is then sent to another module $C_2$ and so on. However, in most cases the processed ciphertext does not preserve the same properties as the fresh one at the first entry, and does not suit exactly what a circuit attends from its predecessor. Therefore, we introduce a novel mechanism called the "Adapter" that allows us to transform and customize an output to a set of gates suiting the input of the next circuit. In particular, when a circuit finishes its processing, and before forwarding the output to the next circuit, it checks how many entries the following circuit is requiring. Then, it uses the "Adapter" to redraw the output to be exactly what the next circuit needs. This way the client's input is processed using multiple modules that form a single chain with several links, see Figure 3.

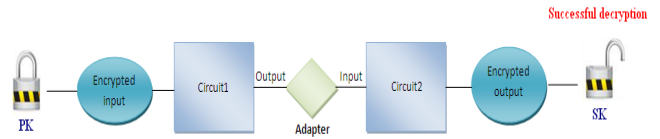

Figure 3. Multi-hop architecture using the Adapter concept.

### C. The Adapter

The use of the Adapter allows us to transform a set of gates to a new sequence while keeping the same functionality. The Adapter enables us to realize a homogeneous multi-hop homomorphic encryption by adapting the output of each circuit to be suitable for the next circuit. The use of the Adapter, however, may violate the privacy since an equivalent part of a circuit is shared with the next circuit. The latter behavior can disclose important circuits' characteristics. To avoid this violation of privacy, we use another type of logical gates called the "Star". The Star, which we proposed in [9], encrypts ordinary gates to hide the operation they implement ("AND" or "XOR"). The Star gate accepts three encrypted bit values, namely, two operands and a flag, which configures the novel operator to be either an "AND" or an "XOR" gate. Thus, a circuit based on the Star gates creates a black box that protects circuit's privacy and makes the "Adapter" technique more secure.

To benefit from the strengths of the Star gate, we propose to view all of the circuits as a set of interrelated Star gates. This approach enables all of these circuits to protect their processing and easily map each circuit's outputs to its next input. This way we manage to realize the Adapter concept without disclosing any sensitive data to the public. As a result, the afore mentioned problem of privacy that is associated with the Adapter concept is overcome.

## V. ENCRYPTED TRUST BASED ROUTING PROTOCOL

In this section we introduce a novel mechanism that computes paths between nodes based on a trust metric and the multi-hop homomorphic mechanism. We assume that a routing algorithm is used to compute the routes prior to the establishment of a trusted path to destination nodes. We further assume that the factors to measure the level of nodes' trustworthiness are already known. In the assumed topology, each node maintains a trust database that includes a summary of all transactions conducted by neighboring nodes. The database associates a trust value with each monitored node.

In our protocol the source node begins by generating both public and secret keys, which are used as bases for the multi-hop homomorphic encryption scheme. Then, the source node executes its trust circuit and calculates the trust value for each neighboring node. The node with the highest trust value is the one to be selected as part of the trusted path. Also, the source node generates a Route Request (RR) packet that carries information about the public key, the destination node and next hop. Moreover, the current element checks the number of inputs required by the next hop and transforms its output using the "Adapter" to a set of encrypted values. These values represent wires that should be used by the next collaborator, where each three values form one Star gate. After that, the source node maps the generated gates to the RR packet, which will also carry the encrypted trust value of the most trusted next hop node. The RR is then forwarded to its next hop to continue discovering the trusted path. Each node receiving the RR firstly verifies whether the destination is among their in-nodes. If the destination is found, the RR is forwarded without making any changes to it. Otherwise, the current node begins by dismantling the received packet and extracting the public key and the encrypted accumulated trust value. Thereafter, the node retrieves the set of gates sent by the previous node, and links them to its circuit to continue calculating the trust path. Then, it changes the contents of the RR by mapping a trusted next hop. Also, the node updates the accumulated trust value for the constructed path as well as the output of "Adapter" mechanism. The RR is then transmitted to the next node. The protocol continues with this procedure until reaching the destination node. Upon receiving the RR packet, the destination replies back with a Route Reply (RP) packet to the source. The RP carries both the whole path from source to destination, besides the encrypted trust value associated with that path. The source node is the only entity that is able to decrypt the received trust value, without being aware of the one associated to each arc.

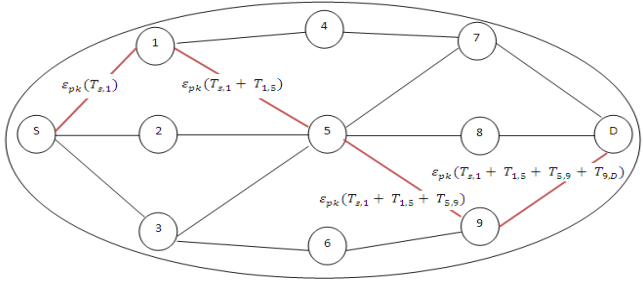

Figure 4. Accumulating encrypted trust value "T" from source to destination.

## VI. PERFORMANCE EVALUATION

In this section we discuss the performance of the proposed protocol. The performance of our system is highly dependent on the parameter of security $\lambda$; the bigger this parameter is the more secure the system becomes. However, as $\lambda$ gets bigger, the system will have to execute more operations.

We study an ad hoc network of 20 nodes that need to select trusted paths to send their packets over. Each node in this topology first needs to evaluate its in-neighbors, and then updates the available trust values before forwarding the packet. This trust value, which varies from 1 (lowest trust level) to 10 (highest trust level), can be represented in 4 bits. The encrypted form of this value is at the order of $4 \times \lambda^5$ bits. Each node participating in the protocol should perform 20 additions and 8 multiplications (homomorphic operations over encrypted values) to update the previous trust value. This means, in order to calculate a path in a network of 20 nodes, we need to compute at most 400 additions and 160 multiplications. To guarantee faster computations, we rely on Karatsuba algorithm [16], which is suitable for operations over big integers. This algorithm allows performing more than $10^9$ integer operations in less than one second (tested on a PC with a 6 GB of RAM memory at a security parameter $\lambda = 3$).

We show in Table 1, the time needed to compute an encrypted trust path under different values of $\lambda$, in a network of 20 nodes.

TABLE I. TIME NEEDED TO COMPUTE ENCRYPTED TRUST PATHS

| Security Parameter $\lambda$ | Time in Seconds |
|---|---|
| 3 | 0.3 |
| 5 | 1 |
| 8 | 3 |
| 10 | 8 |

## VII. CONCLUSION AND PERSPECTIVES

In this paper, we aimed at developing a secure trust-based routing protocol using the concept of multi-hop homomorphic encryption. We propose a model that realizes multi-hop encryption in an environment using one encryption layer. This model is used to realize a secure routing protocol that computes trusted paths based on the behavior of neighboring nodes over a period of time. Our protocol uses a fully homomorphic encryption scheme to protect the privacy of each node while evaluating its trustworthiness.

As a future work, we plan to devise a fully scheme that hides the source and destination nodes while computing the trusted path.

## REFERENCES

[1] A.A. Pirzada, A. Datta, and C. McDonald, "Trust Based Routing for Ad-Hoc Wireless Networks," In *Proceedings of IEEE International Conference on Networks (ICON'04)*, pp. 326-330, 2004.

[2] R.K. Nekkanti and C.-W. Lee, "Trust Based Adaptive On Demand Ad Hoc Routing Protocol," In *ACM Southeast Regional Conference*, pp. 88-93, 2004.

[3] S.-H. Chou, C.-C. Lo, and C.-C. Huang, "Mitigating routing misbehavior in Dynamic Source Routing protocl using trust-based reputation mechanism for wireless ad-hoc networks," In *Consumer Communications and Networking Conference (CCNC)*, pp. 442-446, 2011.

[4] K. Sanzgiri, B. Dahill, B.N. Levine, C. Shields, and E.M. Belding-Royer, "A secure routing protocol for ad hoc networks," In *Proceedings of the 10th IEEE International Conference on Network Protocols*, pp. 78- 87, 2002.

[5] C. Zhang, X. Zhu, Y. Song, and Y. Fang, "A Formal Study of Trust-Based Routing in Wireless Ad Hoc Networks," In *Proceedings of INFOCOM*, pp. 1-9, 2010.

[6] C. Gentry, S. Halevi, and V. Vaikuntanathan, "i-hop homomorphic encryption and rerandomizable Yao circuits," In *CRYPTO*, pp. 155-172, 2010.

[7] T. Schneider, "Practical secure function evaluation," Master's thesis, University of Erlangen-Nuremberg, February 27, 2008.

[8] A.C. Yao, "Protocols for secure computations," In *Proceedings of the 23rd IEEE Sympositum On Foundations of Computer Science*, pp. 160-164, 1982.

[9] Y. Gahi, M. Guennoun, Z. Guennoun, and K. El-khatib, "Encrypted Processes for Oblivious Data Retrieval", In *the 6th International Conference for Internet Technology and Secured Transactions*, pp. 514-518, 2011.

[10] Y. Gahi , M. Guennoun, and K. El-khatib, "A Secure Database System using Homomorphic Encryption Schemes," In *the 3rd Int. Conf. on Advances in Databases, Knowledge, and Data Applications*, pp. 54-58, 2011.

[11] C. Gentry, "A fully homomorphic encryption scheme," PhD thesis, Stanford University, 2009.

[12] D. Boneh, E.-J. Goh, and K. Nissim, "Evaluating 2-DNF formulas on ciphertexts," In *Theory of Cryptography TCC'05*, pp. 325-341, 2005.

[13] P. Paillier,"Public-key cryptosystems based on composite degree residuosity classes," In *EUROCRYPT*, pp. 223-238, 1999.

[14] S. Goldwasser and S. Micali, "Probabilistic encryption and how to play mental poker keeping secret all partial information," In *Proceedings of the fourteenth annual ACM symposium on Theory of computing STOC' 82*, pp. 365-377, 1982.

[15] T. El-Gamal, "A public key cryptosystem and a signature scheme based on discrete logarithms," In *CRYPTO*, pp. 10-18, 1984.

[16] http://en.wikipedia.org/wiki/Karatsuba_algorithm